\documentstyle[12pt]{article}
\def\abstracts#1#2#3{{
        \centering{\begin{minipage}{4.62in}\baselineskip=13pt
        \small
        \centerline{\bf Abstract}
        \vspace*{0.2cm}                
        \parindent=0pt #1\par
        \parindent=18pt #2\par
        \parindent=15pt #3
        \end{minipage} }\par}}
\begin{document}
\vspace*{-2cm}
\hfill \parbox{5cm}{ Mainz preprint\\
                     June 1994}\\
\vspace*{2cm}
\centerline{\LARGE \bf Multibondic Cluster Algorithm for }\\[0.3cm]
\centerline{\LARGE \bf Monte Carlo Simulations of }\\[0.4cm]
\centerline{\LARGE \bf First-Order Phase Transitions}\\[0.6cm]
\vspace*{0.2cm}
\centerline{\large {\em Wolfhard Janke\/} and
                   {\em Stefan Kappler\/}}\\[0.4cm]
\centerline{\large {\small Institut f\"ur Physik,
                    Johannes Gutenberg-Universit\"at Mainz}}
\centerline{    {\small Staudinger Weg 7, 55099 Mainz, Germany }}\\[0.5cm]
\abstracts{}{
Inspired by the multicanonical approach to simulations of first-order
phase transitions we propose for $q$-state Potts models a combination of 
cluster updates with reweighting of the bond configurations in the
Fortuin-Kastelein-Swendsen-Wang representation of this model. 
Numerical tests for the two-dimensional models with $q=7, 10$ and $20$ show
that the autocorrelation times of this algorithm grow with the system size $V$
as $\tau \propto V^\alpha$, where the exponent takes the optimal random walk
value of $\alpha \approx 1$. 
}{}
\vspace*{3.5cm}

\thispagestyle{empty}
\newpage
\pagenumbering{arabic}
%
                     \section{Introduction}
%
Monte Carlo simulations of first-order phase transitions \cite{gunton} in the 
canonical ensemble are severely hampered by extremely large autocorrelation 
times $\tau \propto \exp(2 \sigma L^{D-1})$ where $\sigma$ is the (reduced) 
interface tension between the coexisting phases and $L^{D-1}$ is the 
cross-section of the system \cite{bil}.
To overcome this problem Berg and Neuhaus \cite{bn1,bn2} have recently
introduced  multicanonical simulations which are based on reweighting ideas
and can, in principle, be combined with any legitimate update algorithm. 
Using {\em local} update algorithms (Metropolis or heat-bath) it has been 
demonstrated in several applications \cite{athens} that the growth of 
autocorrelation times with system size is reduced to a power-law, 
$\tau \propto V^\alpha$ with $\alpha \ge 1$. For the two-dimensional $q$-state 
Potts model values of $\alpha \approx 1.3$ have been reported for 
$q=7$ \cite{ours} and $q=10$ \cite{bn2}. 

Since by construction
the multicanonical energy distribution is constant over the interesting
energy range, invoking a random walk argument, one would expect 
an exponent $\alpha = 1$ for an optimally designed update algorithm.  
The purpose of this note is to present for Potts models a cluster update 
variant of the multicanonical approach which is optimal in this sense. 
Basically the idea is to treat the cluster flips in the first place and to 
reweight the bond degrees of freedom instead of the energy.  
\section{The algorithm}

The basis of cluster update algorithms \cite{sw,wolff} is the equivalence 
of the Potts model
\begin{equation}
Z_{\rm Potts} = \sum_{\{\sigma_i\}} e^{-\beta E}; E = -\sum_{\langle ij \rangle}
\delta_{\sigma_i \sigma_j}; \sigma_i = 1,\dots,q,
\label{eq:zpotts}
\end{equation}
with the Fortuin-Kastelein (FK) and random cluster (RC)  
representations \cite{wu},
\begin{equation}
Z_{\rm Potts} = Z_{\rm FK} = Z_{\rm RC},
\label{eq:Z}
\end{equation}
where
\begin{equation}
Z_{\rm FK} = \sum_{\{\sigma_i\}} \sum_{\{b_{ij}\}} \prod_{\langle ij \rangle}
\left[ p \, \delta_{\sigma_i \sigma_j} \delta_{b_{ij},1} + \delta_{b_{ij},0}  
\right],
\label{eq:zfk}
\end{equation}
and
\begin{equation}
Z_{\rm RC} = \sum_{\{b_{ij}\}} p^{ \sum_{\langle ij \rangle} b_{ij} }
q^{ N_c( \{ b_{ij} \})},
\label{eq:zrc}
\end{equation}
with
\begin{equation}
p = \exp(\beta) - 1.
\label{eq:p}
\end{equation}
Here $b_{ij} = 0$ or $1$ are bond occupation numbers and
$N_c( \{ b_{ij} \})$ denotes the number of connected clusters
(including one-site clusters). According to (\ref{eq:zfk}) a Swendsen-Wang 
cluster update sweep then consists in 1) setting $b_{ij}=0$ if $\sigma_i \ne
\sigma_j$, or assigning values 0 and 1 with relative probability $1:p$ if
$\sigma_i = \sigma_j$, 2) identifying clusters of spins that are connected 
by ``active'' bonds ($b_{ij}=1$), and 3) choosing a new random value 
$1 \dots q$ independently for each cluster. 

By differentiating $\ln Z$ with respect to $\beta$ it is easy to see that
the average of the energy 
$E = - \sum_{\langle ij \rangle} \delta_{\sigma_i \sigma_j}$
can be expressed in terms of the average of the number of active bonds
$B = \sum_{\langle ij \rangle} b_{ij}$,
\begin{equation}
\frac{\partial \ln Z}{\partial \beta} = - \langle E \rangle = \frac{p+1}{p}
\langle B \rangle,
\label{eq:e_b}
\end{equation}
and for the specific heat per site $C$ one finds
\begin{equation}
C V /\beta^2 = -\frac{\partial \langle E \rangle}{\partial \beta} = 
-\frac{p+1}{p^2} \langle B \rangle + \left( \frac{p+1}{p} \right)^2 
\left( \langle B^2 \rangle - \langle B \rangle^2 \right).
\label{eq:c_b}
\end{equation}
Eq.~(\ref{eq:e_b}) suggests that the bond histogram $P_{\rm can}^{\rm bond}(B)$ 
should develop for $\beta = \beta_t \pm \delta \beta$ a pronounced peak around 
$B_{o,d} = - \frac{p}{p+1} E_{o,d}$, and for $\beta \approx \beta_t$ a 
double-peak structure similar to $P_{\rm can}^{\rm ene}(E)$.
In fact, as is illustrated in Fig.~1 for $q=7$ and $L=60$, a plot of 
$P_{\rm can}^{\rm bond}$ versus $\frac{p+1}{p} B$ is hardly distinguishable 
from $P_{\rm can}^{\rm ene}(E)$. For other values of $q$ and $L$ the 
comparison looks very similar.

In terms of $P_{\rm can}^{\rm bond}$ the slowing down of canonical
simulations is thus caused by the strongly suppressed configurations near the
minimum between the two peaks, analogous to the well-known argument for 
$P_{\rm can}^{\rm ene}$. To enhance these probabilities we therefore introduce 
in analogy to multicanonical simulations a ``multibondic'' partition function
\begin{equation}
Z_{\rm mubo} = \sum_{\{\sigma_i\}} \sum_{\{b_{ij}\}} \prod_{\langle ij \rangle}
\left[ p \, \delta_{\sigma_i \sigma_j} \delta_{b_{ij},1} + \delta_{b_{ij},0}
\right] \exp(-f_{\rm bond}(B)),
\label{eq:zmubo}
\end{equation}
where $f_{\rm bond}(B) = \ln P_{\rm can}^{\rm bond}(B)$ between the two peaks 
and $f_{\rm bond}(B) = 0$ otherwise. Of course, as in multicanonical 
simulations, any reasonable approximation of $P_{\rm can}^{\rm bond}(B)$ 
can be used in practice.
Canonical expectation values can always be recovered exactly by applying the 
inverse of the reweighting factor $\exp(-f_{\rm bond}(B))$.

Obviously, once the $b_{ij}$ are given, we can update the $\sigma_i$ exactly 
as in the original Swendsen-Wang cluster algorithm.
To update the $b_{ij}$ we proceed as follows. If $\sigma_i \ne \sigma_j$
then the bond $b_{ij}$ is never active and we always set $b_{ij}=0$. If
$\sigma_i = \sigma_j$ then we define $B' = B - b_{ij}$ and choose new values
$b_{ij} = 0$ or $1$ with probabilities 
${\cal P}(b_{ij}=0) = {\cal N} \exp(-f_{\rm bond}(B'))$
and ${\cal P}(b_{ij}=1) = {\cal N} p \, \exp(-f_{\rm bond}(B'+1))$, 
where ${\cal N}$ is a trivial normalization factor
(${\cal N} = 1/\left[ \exp(-f_{\rm bond}(B')) + 
p \, \exp(-f_{\rm bond}(B'+1))\right]$ ).
Since this is nothing but a local heat-bath algorithm for the $b_{ij}$ the
whole procedure is obviously a valid update algorithm.
\section{Results}

To evaluate the performance of the multibondic (mubo) cluster algorithm we 
performed simulations of the Potts model (\ref{eq:zpotts}) in two dimensions 
with $q = 7, 10$ and $20$. The investigated lattice sizes and simulation 
temperatures are compiled in Table~1. For comparison we run for the same 
parameters also standard multicanonical (muca) simulations using the heat-bath 
update algorithm. In each run we recorded $N = 100\,000$
measurements of $E$, $B$ and two definitions of the magnetization in a 
time-series file. (The only exception is the multicanonical simulation
for $q=7$, $L=100$ with $N=30\,000$.) Between the
measurements we performed $M$ lattice sweeps, with $M$ adjusted in such a way
that the autocorrelation times in units of measurements and thus the effective 
statistics of practically uncorrelated data was roughly the same in all 
simulations. 

To make sure that the new multibondic algorithm was implemented correctly, we 
have analyzed some of the usually considered canonical quantities such as the
specific heat $C$ and the Binder parameter 
$V = 1 - \langle E^4 \rangle/3 \langle E^2 \rangle^2$.
In Table~2 we compare results
for the specific-heat maximum and Binder-parameter minimum for $q=10$ obtained
from our multicanonical and multibondic simulations. The error bars are
estimated by the jackknife method \cite{jackknife}. The data are in good 
agreement with each other and also with results from independent canonical 
Metropolis \cite{blm1} or single-cluster \cite{icke} high-statistics 
simulations.

As discussed in Refs.\cite{js_letter,js_paper} in the context of a 
multicanonical multigrid implementation it is not completely obvious
which definition of the autocorrelation time should be used to characterize the
dynamics of multicanonical or multibondic simulations. One could, e.g., analyze
the (multicanonical/bondic) autocorrelation function of $E$, 
$E \exp(f_{\rm ene}(E))$, $B$, or $B \exp(f_{\rm bond}(B))$, where 
$f_{\rm ene}(E)$ is the multicanonical analogue
of $f_{\rm bond}(B)$. More relevant from a practical point of
view is the effective autocorrelation time \cite{js_letter,js_paper} 
for canonical observables which
can be defined from the ratio of proper and naive error estimates 
($\epsilon/\epsilon_{\rm naiv} = \sqrt{2 \tau^{\rm eff}}$).
The third possibility, which allows a direct comparison with previous work,
is to define flipping (or more properly diffusion) times 
$4 \tau^{\rm flip}_E$ by
counting the number of update sweeps that are needed to travel from 
 $E < E_{\rm min}$ to $E > E_{\rm max}$ and back. Here $E_{\rm min,max}$
(or $B_{\rm min,max}$ for $\tau^{\rm flip}_B$) are cuts which are usually
chosen as the peak locations $E_{o,d}(L)$ of the canonical probability
distribution. Alternatively one could also use the infinite volume limits
$\hat{E}_{o,d}$ of $E_{o,d}(L)$ for all lattice sizes. For 2D Potts models
this is straightforward since $\hat{E}_{o,d}$ are known exactly. In our
simulations we have tested if $E$ (or $B$) has passed the cuts after each 
sweep. We observed significantly larger $\tau^{\rm flip}$ when performing 
this test only every $M$'th sweeps, since then any cut-crossings during the 
$M-1$ sweeps between the measurements cannot be detected.

Our results for $\tau^{\rm flip}_E$ obtained in multicanonical and multibondic 
simulations for $q=7, 10$ and $20$ are shown in the log-log plots of Figs.~2-4.
Here we have used the canonical peak locations for the energy cuts. 
Let us first concentrate on the results for $q=7$ in Fig.~2 where we
have included for comparison the data from previous multicanonical 
simulations \cite{ours} and also Rummukainen's results for his 
hybrid-like two-step algorithm which combines microcanonical cluster updates
with a multicanonical demon refresh \cite{ru2}. Both cluster update 
versions show qualitatively the same behavior and, for $L > 20$, perform
much better than the standard multicanonical algorithm. From least-square fits
to
\begin{equation}
\tau^{\rm flip}_E = a V^{\alpha}
\label{eq:tau}
\end{equation}
we estimate $\alpha \approx 1.3$ for multicanonical heat-bath and 
$\alpha \approx 1$ for multibondic cluster simulations; see Table~3
where we also give results for fixed energy cuts.  
Our results for $\tau^{\rm flip}_B$ are almost indistinguishable from
$\tau^{\rm flip}_E$ which, recalling Fig.~1, is no surprise. 
Furthermore we have also measured the effective autocorrelation times 
and find that they are systematically smaller for both algorithms. 

Unfortunately, for $q=10$ and $20$ the situation is less favorable for the
multibondic algorithm. While we still find an exponent of $\alpha \approx 1$,
the prefactor in (\ref{eq:tau}) turns out to be so large that we can take 
advantage of this asymptotic improvement only for very large lattice
sizes. As can be seen in Fig.~3, for $q=10$ the cross-over happens around
$L=50$. Extrapolating to $L=100$ we estimate that the multibondic algorithm
would perform for this lattice size about 1.5 times faster than the 
standard multicanonical heat-bath. For $q=20$ the same comparison clearly 
favors the standard algorithm for all reasonable lattice sizes 
- and we certainly cannot recommend the new algorithm for large $q$.
\section{Conclusions}

In summary, we have proposed for Potts models a combination of cluster 
update techniques 
with reweighting in the random bond representation
and shown that this approach is feasible in practice. In fact, it
is technically not more involved than the standard multicanonical approach
and one lattice sweep takes about the same CPU time. 
Numerical tests for the two-dimensional $q$-state Potts model with $q=7,10$ 
and $20$ show that
the multibondic cluster algorithm is optimal in the sense that the exponent
$\alpha$ in the power-law, $\tau^{\rm flip} = a V^\alpha$,
is consistent with $\alpha = 1$, the value one would expect in an
idealized random walk picture. For $q=7$ the multibondic algorithm clearly
outperforms the standard multicanonical heat-bath algorithm. Compared with
Rummukainen's hybrid-like two-step cluster variant the multibondic 
autocorrelation times are smaller for all lattice sizes by a roughly constant 
factor of $1.5$.
For larger values of $q$, however, the prefactor $a$ turns out to be 
relatively large, rendering the new algorithm for reasonable lattice sizes more 
efficient than multicanonical simulations only for $q < q_0$ with $q_0$ 
somewhat above 10. 

The multibondic cluster algorithm may be of value for a wide range of 
investigations, since it can be applied to any systems where conventional 
cluster update techniques are applicable.
\section*{ACKNOWLEDGEMENTS}

W.J. would like to thank the DFG for a Heisenberg fellowship and 
S.K. gratefully acknowledges a fellowship by the Graduiertenkolleg 
``Physik und Chemie supramolekularer Systeme''. The Monte Carlo simulations 
were performed on the CRAY Y-MP of the H\"ochstleistungsrechenzentrum J\"ulich,
the CRAY Y-MP's of the Norddeutscher Rechnerverbund in Kiel and Berlin under
grant bvpf03, and on the Mainz cluster of fast RISC workstations.
%

%
%
\newpage
%
%
\begin{table}              
{\Large\bf Tables}\\[1cm]
 \begin{center}
  \caption[a]{\label{tab:statistic} Simulation parameters: $\beta_0$ is the
  inverse simulation temperature, $M_{\rm muca}$ and $M_{\rm mubo}$ denote
  the number of lattice sweeps between measurements, and $|E|_{\rm min,max}$
  and $B_{\rm min,max}$ are the cuts used in the definition of the 
  autocorrelation time $\tau^{\rm flip}$.\\}
  \begin{tabular}{|r|r|r|r|r|r|r|r|r|}
\hline
\multicolumn{1}{|c|}{$q$}                 &
\multicolumn{1}{c|}{$L$}                  &
\multicolumn{1}{c|}{$\beta_0$}            & 
\multicolumn{1}{c|}{$M_{\rm muca}$}       &
\multicolumn{1}{c|}{$|E|_{\rm min}$}      &
\multicolumn{1}{c|}{$|E|_{\rm max}$}      &
\multicolumn{1}{c|}{$M_{\rm mubo}$}       &
\multicolumn{1}{c|}{$B_{\rm min}$}        &
\multicolumn{1}{c|}{$B_{\rm max}$}        \\
\hline
   &  20 & 1.284690 &  10 &   426 &   644 &  10 &  310 &   462 \\
 7 &  40 & 1.291050 &  50 &  1801 &  2542 &  30 & 1310 &  1840 \\
   &  60 & 1.292283 & 100 &  4139 &  5675 &  70 & 3003 &  4118 \\
   & 100 & 1.293089 & 100 & 11692 & 15672 & 100 & 8476 & 11367 \\
\hline
    & 12 & 1.407380 &  10 &  120 &  247 &  20 &   91 &  186 \\
    & 16 & 1.415340 &  20 &  222 &  439 &  40 &  166 &  326 \\
10  & 20 & 1.418864 &  30 &  353 &  676 &  60 &  270 &  512 \\
    & 26 & 1.421642 &  70 &  614 & 1137 & 140 &  467 &  867 \\
    & 34 & 1.423380 & 200 & 1065 & 1938 & 400 &  813 & 1474 \\
    & 50 & 1.424752 & 200 & 2349 & 4185 & 400 & 1797 & 3182 \\
\hline
    &  4 & 1.577747 &   5 &   6 &  32 &  10 &   5 &  24 \\
    &  6 & 1.639809 &   6 &  17 &  72 &  12 &  14 &  56 \\
    &  8 & 1.665033 &  12 &  33 & 124 &  25 &  28 &  96 \\   
    & 10 & 1.676647 &  17 &  56 & 192 &  35 &  45 & 151 \\
20  & 12 & 1.683517 &  32 &  79 & 280 &  65 &  66 & 215 \\
    & 14 & 1.688195 &  67 & 112 & 357 & 135 &  92 & 295 \\
    & 16 & 1.690278 &  87 & 154 & 470 & 175 & 122 & 384 \\
    & 18 & 1.692013 & 125 & 194 & 593 & 250 & 155 & 485 \\
    & 20 & 1.693698 & 175 & 231 & 728 & 350 & 201 & 595 \\
\hline
\end{tabular}
\end{center}
\end{table}
%
%
\begin{table}[htb]
\caption[a]{\label{tab:test} Two-dimensional 10-state Potts model: Comparison 
of results for specific-heat maxima and Binder-parameter minima from 
multicanonical (muca) and multibondic (mubo) simulations.}
\begin{center}
\begin{tabular}{|r|r|l|r|l|r|}
\hline
 \multicolumn{1}{|c}{$L$}                     &
 \multicolumn{1}{|c}{alg.}                  &
 \multicolumn{1}{|c}{$\beta_{C_{\rm max}}$} &
 \multicolumn{1}{|c}{$C_{\rm max}$}         &
 \multicolumn{1}{|c}{$\beta_{V_{\rm min}}$} &
 \multicolumn{1}{|c|}{$V_{\rm min}$}         \\
\hline
12 & muca & 1.40621(28)  &  44.89(18)  & 1.39256(29)  & 0.50379(75) \\
12 & mubo & 1.40733(29)  &  44.92(18)  & 1.39373(29)  & 0.50521(79) \\
\hline
16 & muca & 1.41480(17)  &  73.68(25)  & 1.40757(17)  & 0.52503(53) \\
16 & mubo & 1.41451(15)  &  73.87(27)  & 1.40729(15)  & 0.52447(61) \\
\hline
20 & muca & 1.41837(11)  & 109.92(32)  & 1.41392(11)  & 0.53541(44) \\
20 & mubo & 1.418602(95) & 109.33(33)  & 1.414150(97) & 0.53649(44) \\
\hline
26 & muca & 1.421325(71) & 177.61(44)  & 1.418786(72) & 0.54451(34) \\
26 & mubo & 1.421506(54) & 177.91(43)  & 1.418969(54) & 0.54453(34) \\
\hline
34 & muca & 1.423313(36) & 296.13(52)  & 1.421873(36) & 0.55001(24) \\
34 & mubo & 1.423312(26) & 297.14(50)  & 1.421872(26) & 0.54941(23) \\
\hline
50 & muca & 1.424801(28) & 627.79(97)  & 1.424155(34) & 0.55437(21) \\
50 & mubo & 1.424834(21) & 627.6(1.1)  & 1.424188(21) & 0.55439(22) \\
\hline
\end{tabular}
\end{center}
\end{table}
%
%
\begin{table}[htb]
\caption[a]{\label{tab:zvalues}Results for the dynamical exponent $\alpha$
in multicanonical (muca) and multibondic (mubo) simulations from fits to
$\tau^{\rm flip}_E = a V^\alpha$, using $L$-dependent cuts
defined by the canonical peak locations
($\alpha_{\rm max}$) and fixed cuts at the infinite volume limits of
$E_{d,o}$ and $B_{d,o}$ ($\alpha_{\rm fix}$).}
\begin{center}
\begin{tabular}{|r|l|l|l|l|}
\hline
                                           &
 \multicolumn{2}{c|}{$\alpha_{\rm max}$}   &
 \multicolumn{2}{c|}{$\alpha_{\rm fix}$}   \\
\hline
 \multicolumn{1}{|c|}{$q$}                 &
 \multicolumn{1}{c|}{muca}                 &
 \multicolumn{1}{c|}{mubo}                 &
 \multicolumn{1}{c|}{muca}                 &
 \multicolumn{1}{c|}{mubo}                 \\
\hline
  7 & 1.27(2)   & 0.92(2)   & 1.53(2)   & 1.02(2) \\
 10 & 1.32(2)   & 1.05(1)   & 1.43(1)   & 1.12(1) \\
 20 & 1.26(1)   & 1.09(1)   & 1.46(1)   & 1.18(1) \\
\hline
\end{tabular}
\end{center}
\end{table}
%
%
\clearpage
\newpage
  {\Large\bf Figure Headings}
  \vspace{1in}
  \begin{description}
    \item[\tt\bf Fig. 1:]
Canonical energy and bond histograms for $q=7$, $L=60$, and 
$\beta=1.292283$. The bond histogram is plotted vs $[(p+1)/p]B$, where
$p =\exp(\beta)-1$.
    \item[\tt\bf Fig. 2:]
Log-log plot of autocorrelation times $\tau^{\rm flip}_E$ of the energy 
vs lattice size for $q=7$, using $L$-dependent energy cuts defined by the
peak locations of the canonical energy distribution.
    \item[\tt\bf Fig. 3:]
Same as Fig.~2 for $q=10$.
    \item[\tt\bf Fig. 4:]
Same as Fig.~2 for $q=20$.
  \end{description}
\newpage

%
%
\begin{figure}[t]
\vskip 8.0truecm
\includegraphics{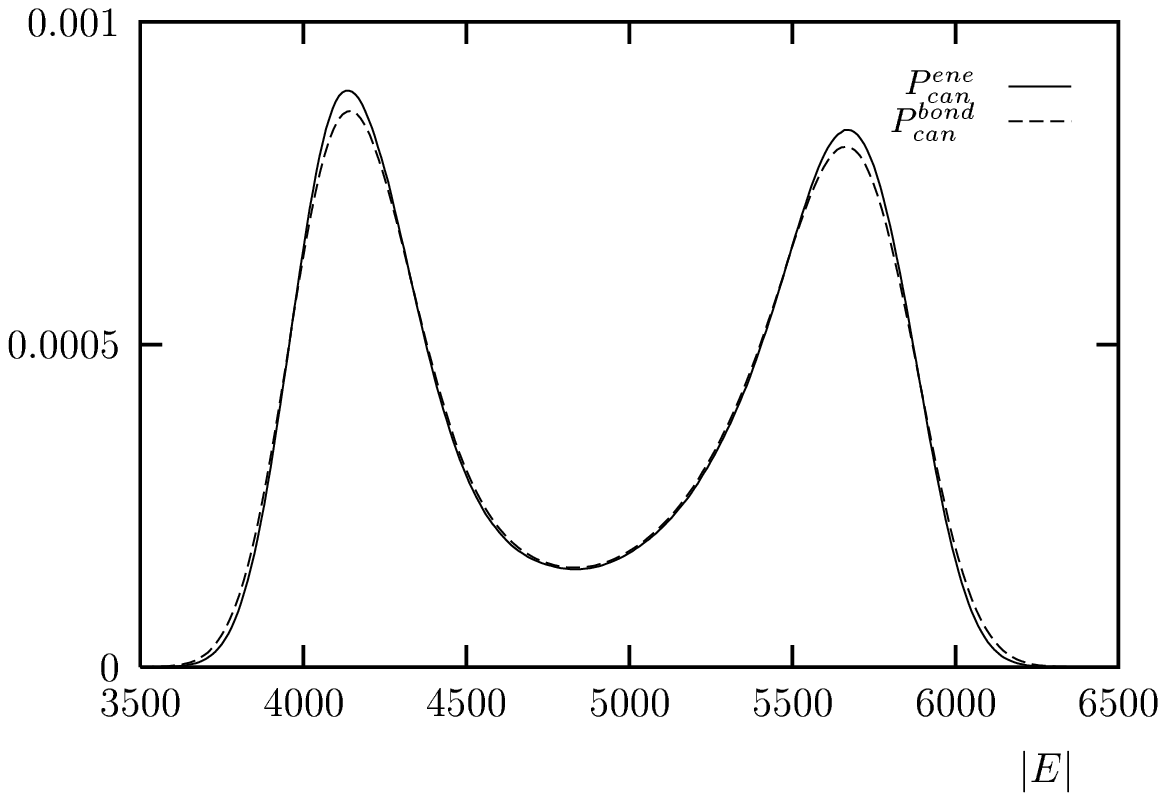}
\caption{\label{fig1}}
\end{figure}
\begin{figure}[b]
\vskip 8.0truecm
\includegraphics{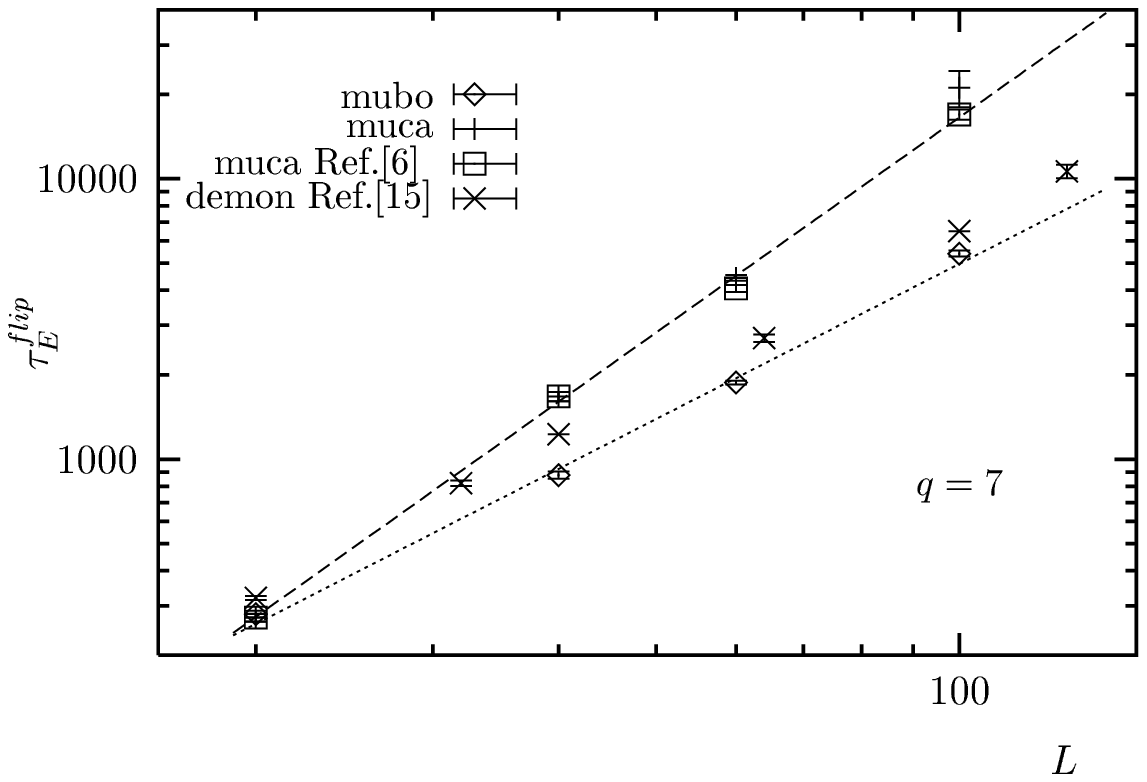}
\caption{\label{fig2}}
\end{figure}
\newpage
\begin{figure}[t]
\vskip 8.0truecm
\includegraphics{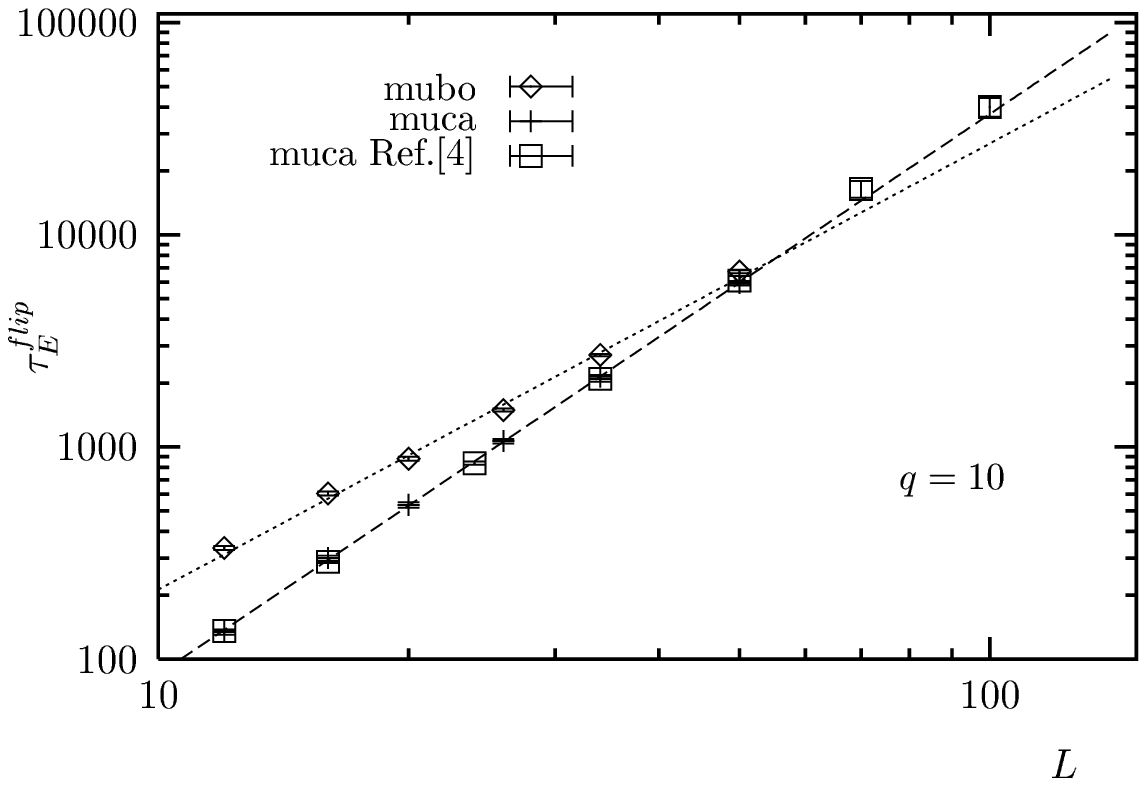}
\caption{\label{fig3}}
\end{figure}
\begin{figure}[b]
\vskip 8.0truecm
\includegraphics{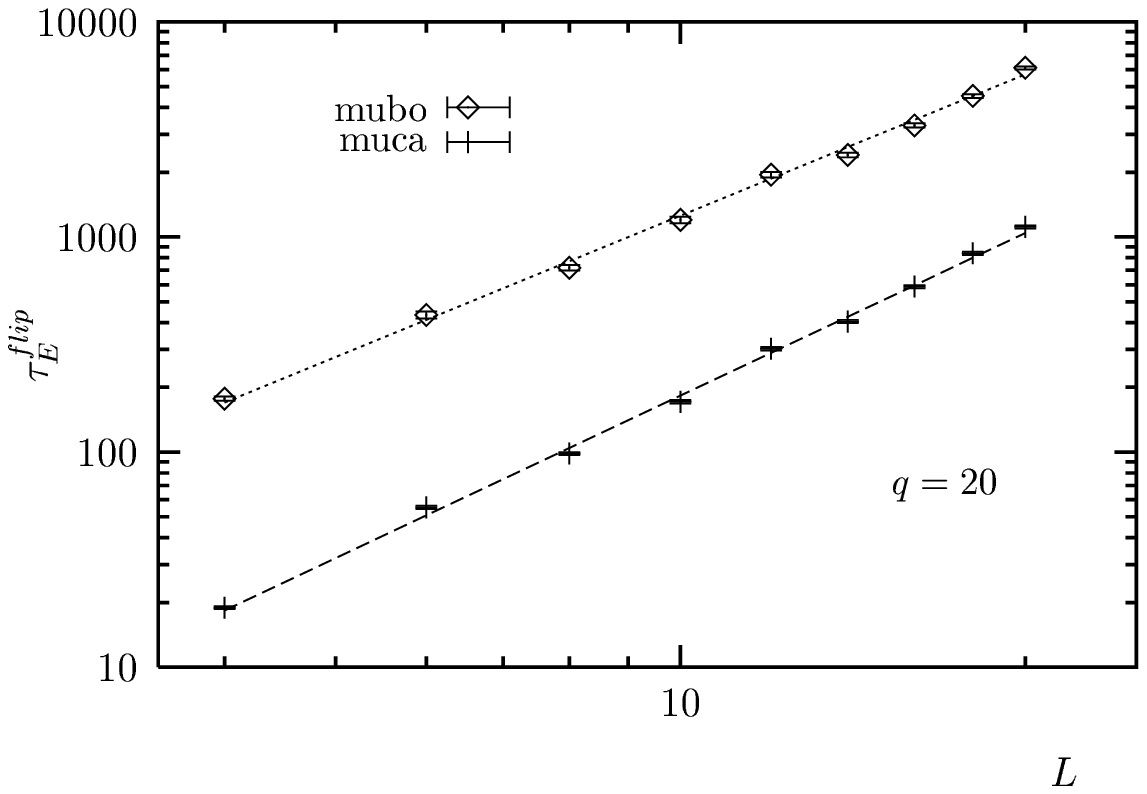}
\caption{\label{fig4}}
\end{figure}
\end{document}